\documentclass[12pt]{amsart}

\usepackage{graphicx}
\usepackage{psfrag} 
\usepackage{mathptmx}      
\usepackage{amssymb}
\usepackage{amsmath}
\usepackage{amsthm}
\usepackage[dvipsnames]{xcolor}
\usepackage{algorithm}
\usepackage{marvosym}
\usepackage{a4wide}
\usepackage[noend]{algpseudocode}
\algrenewcommand\algorithmicrequire{\textbf{Input:}}
\algrenewcommand\algorithmicensure{\textbf{Output:}}

\usepackage{color}

\newcommand{\leo}[1]{#1}

\usepackage[mathscr]{euscript}

\newtheorem{theorem}{Theorem}[section]
\newtheorem{lemma}[theorem]{Lemma} 

\newtheorem{proposition}[theorem]{Proposition}

\title{Is this network forest-based?}
\author{K.T. Huber, V. Moulton, G. E. Scholz, L. van Iersel}
\thanks{School of Computing Sciences, 
	University of East Anglia, UK}

\date{\today}

\begin{document}

\begin{abstract}
In evolutionary biology, networks are becoming 
increasingly used to represent evolutionary histories for
species that have undergone non-treelike or reticulate evolution.
Such networks are essentially directed acyclic graphs with a leaf set that corresponds to 
a collection of species, and in which non-leaf vertices with indegree 1 correspond to 
speciation events and vertices with indegree greater than 1 correspond to 
reticulate events such as  gene transfer. 
Recently forest-based networks have been
introduced, which are essentially (multi-rooted) networks that can be formed
by adding some arcs to a collection of phylogenetic trees (or phylogenetic forest),
where each arc is added in such a way that its ends always lie in
two different trees in the forest. In this paper, we consider 
the complexity of deciding whether or not a given network is
proper forest-based, that is, whether it can be formed by adding arcs to some underlying phylogenetic forest 
which contains the same number of trees as there are roots in the network. More specifically,
we show that it can be decided in polynomial time whether or not a binary, tree-child network with $m \ge 2$ roots is proper forest-based in case $m=2$, but 
that this problem is NP-complete for $m\ge 3$. We also 
give a fixed parameter tractable (FPT) algorithm for deciding 
whether or not a network in which every vertex has indegree at most 2 is proper forest-based.
A key element in proving our results is a new characterization for when 
a network with $m$ roots is proper forest-based 
which is given in terms of the existence of certain $m$-colorings 
of the vertices of the network.\\

\end{abstract}

\maketitle

\section{Introduction}

\begin{figure}[h]
	\begin{center}
		\includegraphics[scale=0.7]{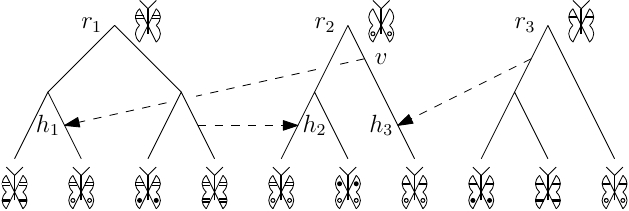}
	\end{center}
	\caption{A proper forest-based network $N$ on ten leaves.
 Each of the three phylogenetic trees in the underlying forest represents a hypothetical 
 butterfly lineage with main wing pattern indicated
 next to the root of the tree.  The network $N$ is the result of adding dashed arcs in between 
 pairs of trees
 in the forest. Each added arc  
 corresponds to some genetic material being introduced into a lineage from one
 of the others, which results in a wing pattern change for the descendants.}
 \label{fig:intro}
\end{figure}

Recently, the concept of forest-based networks has been introduced 
within the area of phylogenetics \cite{HMS22a}. 
Informally, a forest-based network is defined as follows (full
definitions of the terms used in the introduction are given in
the next section). A {\em phylogenetic tree} is 
a rooted tree whose leaf-set corresponds to a collection of taxa or species; a 
{\em phylogenetic forest} is a collection of leaf-disjoint phylogenetic trees. 
A {\em forest-based network} is essentially a directed acyclic graph
$N$ that can be formed by adding a set of arcs to 
a phylogenetic forest so that the end vertices 
of each added arc lie in two different trees of the forest; $N$ is {\em proper} 
if the number of sources or roots of the network is 
equal to the number of trees in the forest. For 
example, the network in Figure~\ref{fig:intro} is 
proper forest-based. Forest-based networks can be regarded as a certain type of 
{\em phylogenetic network}, and are related to 
the intensively studied {\em tree-based networks}
\cite{francis2015phylogenetic} (see e.g.  \cite{S16,kong2022classes} for  
recent reviews of phylogenetic networks, including more details
concerning tree-based networks). 

Forest-based networks arise in the study of reticulate evolutionary processes in 
which species exchange genetic information through processes such 
as introgression \cite{SPMMH19} and lateral gene transfer \cite{HMS22a}.
In the case of introgression, the phylogenetic trees underlying 
a forest-based network correspond to the evolutionary 
histories of different subgroups or lineages within a certain species.
The arcs in between different trees then represent the past interchange of 
genetic material between these lineages. A well-studied example of this phenomenon 
is
butterfly evolution, where the genetic material that is swapped 
between lineages influences wing patterns \cite{edelman2019genomic,wallbank2016evolutionary}.
Figure~\ref{fig:intro} shows a hypothetical example to illustrate this phenomenom.
The use of a special type of forest-based network called an 
{\em overlaid species forest} to analyse introgression can
be found in~\cite{SPMMH19}.

In this paper we are interested in the following recognition problem
for {\em networks} (a certain type of directed acyclic graph
as defined in the next section):

\begin{itemize}
\item[(P)] Is a given network $N$ proper forest-based? 
\end{itemize}

Our main results are as follows. A network 
is {\em binary} if \leo{all vertices have indegree and outdegree at most~$2$ and all non-root vertices have overall degree~$1$ or~$3$.}
A network is \emph{tree-child} if each non-leaf vertex has at least one child with indegree 1.
For a network 
$N$ with~$v$ vertices, $m$ roots, $n$ leaves, $r$ vertices 
with indegree at least~$2$, \leo{and maximum outdegree~$\Delta$,} we shall prove that:

\begin{itemize}
    \item[(R1)] If~$N$ is binary, tree-child and $m=2$, then (P) can be decided
    in $O(n\cdot r)$ time.
    \item[(R2)] If~$N$ is binary, tree-child and $m\geq 3$, then  (P) is NP-complete.
    \item[(R3)] If~$N$ is tree-child, $m\ge 2$, with maximum outdegree~$2$ and 
    maximum indegree at most~$3$, then (P) is NP-complete.
    \item[(R4)] If every vertex in $N$ has indegree at most 2, then there is a fixed parameter 
    tractable algorithm with parameters~$r,m$ \leo{and~$\Delta$} for deciding (P), which is 
    linear in~$v$.
\end{itemize}

We now briefly summarise the rest of the contents of this paper.
In the next section, we present some formal 
definitions concerning networks.
In Section~\ref{sec:char}, we derive a key characterization for when a
network with $m \ge 2$ roots is proper forest-based in terms of 
the existence of an $m$-coloring of the vertices of the network that satisfies certain
properties (Lemma~\ref{lm-fbcol}).
In Section~\ref{sec:quasi-binary}, we establish Statement~(R3)
by reducing from the \textsc{Set-Splitting} problem (Theorem~\ref{pr-2nph}).
Using colorings, in Section~\ref{sec:alternative} we present an alternative proof for Statement~(R3) 
in the special case $m\ge 3$ by reducing from the \textsc{Graph $m$-colorabilty} problem.  
The construction that we use in this proof is then used again 
in Section~\ref{sec:binary} to prove that Statement~(R2) 
holds (Theorem~\ref{pr-3nph-binary}). 
Using the concept of so-called omni-extensions  \cite{HMS22a} 
we also prove Statement~(R1) (Theorem~\ref{pr-binary-linear}). Finally,  
in Section~\ref{sec:ftp-alg} we prove Statement~(R4), before concluding
in Section~\ref{sec:conclusion} with a brief discussion of 
potential directions for future research.

\section{Definitions}

From now on, $X$ is a finite set with $|X|\geq 2$.
Suppose that $N$ is a connected, directed acyclic graph (DAG) and~$v$ is 
a vertex in $N$. Then \leo{$v$ is a {\em leaf} if it is a sink vertex},
$v$ is a {\em root} if it is a source vertex, and  
$v$ is a {\em reticulation} if it has indegree at least 2. If $v$ is not a leaf, then $v$ \leo{is} an {\em internal vertex} of $N$. 
We call an arc $a=(u,v)$ of $N$ {\em internal} if $u$ and $v$ are both internal vertices of $N$. 
If $v$ is an internal vertex that has indegree~$1$ or less, then $v$ is a {\em tree-vertex}.

We call $N$ a {\em network (on $X$)} if it has leaf-set $X$ (which we also denote by $L(N)$), \leo{each leaf has indegree~$1$},
every root has outdegree at least 2, 
every reticulation has outdegree 1, and there is no vertex with indegree and outdegree 1. If $N$ has $m$ roots, we  also 
call it an {\em $m$-network}\footnote{Note that this is more general
than the definition of an $m$-network given in \cite{HMS22a}.}.
We say that two $m$-networks $N$ and $N'$ on $X$ 
are {\em isomorphic} if there is a DAG isomorphism between
$N$ and $N'$ which is the identity on $X$. 

\leo{A network is \emph{semi-binary} if all reticulations have indegree~$2$. It }
is \emph{quasi-binary} if all tree-vertices have outdegree $2$ 
and all reticulations have indegree~$2$ or~$3$. 
\leo{Note that a network is binary if it is quasi-binary and semi-binary}. A network is \emph{tree-child} if every internal
vertex has a child that is a tree-vertex or a leaf.

A {\em phylogenetic network (on $X$)} is a network on $X$ with 
exactly one root and a {\em phylogenetic tree (on $X$)} is  a phylogenetic network on $X$ with no 
reticulations. For technical reasons, we shall also call an isolated vertex 
$v$ a phylogenetic tree (on $\{v\}$).
We say that two distinct leaves $x,y$ of a phylogenetic tree $T$ {\em form  a cherry} 
if they share a parent, and we denote such a cherry by $\{x,y\}$.
We call a phylogenetic tree $T$ on $X$ a {\em caterpillar tree (on $X$)} 
if $T$ has a unique cherry and every internal arc of $T$ lies on the 
directed path from the root of $T$ to the parent of the cherry.

A {\em phylogenetic forest (on $X$)} is a directed graph $F$ whose 
connected components are phylogenetic trees 
and such that $L(T)\cap L(T')=\emptyset$ for any 
distinct trees $T,T'$ in $F$ and $X=\bigcup_{T\in F}L(T)$. 
For convenience, we will sometimes call a non-empty set of phylogenetic trees a phylogenetic forest.

Suppose $N=(V,A)$ is an $m$-network on $X$, for some $m\geq 2$.  Then 
$N$ is called {\em forest-based} if there exists a subset $A'\subseteq A$ 
such that $F'=(V,A')$ is a forest \leo{with leaf set} $X$ and every arc in $A-A'$
has end vertices that are in different trees of $F'$. 
We call $F'$ a {\em subdivision forest} of~$N$. Moreover
the phylogenetic forest~$F$ obtained from~$F'$ by repeatedly 
suppressing any vertices of indegree and outdegree 1 and 
any outdegree 1 roots until this is no longer possible is called a {\em base forest} of~$N$. 
Note that, in particular, we can think of~$F$ as being embedded within~$N$. 
In addition, we call a forest-based $m$-network
{\em proper forest-based} 
if it has a base forest containing~$m$ trees. See \cite{HMS22a,HMS22b} for more on such networks.

 
\section{Colorings and proper forest-based networks}
\label{sec:char}

For $G$ a (simple) graph and $C\not=\emptyset$ a finite set, a surjective map $\sigma: V(G) \to C$ is 
a \emph{$|C|$-coloring
\leo{of $G$}}. We sometimes refer to 
the elements of $C$ as {\em colors}. For $v \in V(G)$, we call  $\sigma(v)$ 
the \emph{color of $v$ under $\sigma$}, or 
simply the {\em color of $v$} if $\sigma$ is clear from the context.
We call a coloring $\sigma$ of $G$ {\em proper} if $\sigma(x)\not=\sigma(y)$
holds for any two adjacent vertices $x,y\in V(G)$. We now prove a useful lemma which characterizes proper forest-based networks
in terms of colorings of their vertex sets.

\begin{lemma}\label{lm-fbcol}
	Let $N$ be an $m$-network on $X$, $m\geq 2$. Then $N$ is proper forest-based if and only 
	if there exists an $m$-coloring of $N$ such that:
	\begin{itemize}
		\item[(C1)] Each non-root vertex of $N$ has the same color as exactly one of its parents.
		\item[(C2)] Each internal vertex of $N$ has the same color as at least one of its children.
	\end{itemize}
\end{lemma}

\begin{proof}
	Suppose first that $N$ is proper forest-based. Let $F'$ be a \leo{corresponding} subdivision forest of $N$.  
 We claim that the 
	map $\sigma: V(N) \to F'$ that assigns to each $v \in V(N)$ the tree in $F'$ that contains $v$ 
	in its vertex set is an $m$-coloring of $N$ that satisfies (C1) and (C2).
	
	To see (C1), let $v $ be a \leo{non-root} vertex of $N$. 
 \leo{Since the root of each tree of~$F'$ is a root of~$N$,}~$v$ 
	is not the root of $\sigma(v)$. So at least one of the parents of $v$ has the same color 
 \leo{as $v$}. If $v$ has two or more parents enjoying this property, then $v$ 
	has indegree $2$ or more in $\sigma(v)$, a contradiction as $\sigma(v)$ is a tree. Hence, (C1) holds.
	
	To see (C2), let $v$ be \leo{an internal} vertex of $N$. 
 Since 
 $L(N)=L(F')$, 
 \leo{$v$ is not} a leaf of $\sigma(v)$. 
	Hence, there exists at least one child $u$ of $v$ in $N$ such that $\sigma(u)=\sigma(v)$. 
	Thus, (C2) holds. 
	
	Conversely, suppose that there exists a coloring $\sigma$ of $N=(V,A)$ satisfying (C1) and (C2). 
	Let $F'$ be the graph with vertex set $V$ and edge set
	$A'=\{(u,v) \in A\,:\, \sigma(u)=\sigma(v)\}$. By (C1), no vertex in $F'$ 
	has indegree greater than $1$. So, $F'=(V, A')$ is a subdivision of $N$ in the form of a forest. 
	By (C2), we have  $L(N)=L(F')$ and so $F'$ is a subdivision forest of $N$. \leo{Since the end vertices of each arc in $A-A'$ have different colors, they appear in different connected components (trees) of~$F'$. Hence,~$N$ is proper forest-based.}
\end{proof}

Note that if $\sigma$ is an $m$-coloring of an $m$-network $N$ satisfying (C1) and (C2), then 
these two properties, together with the fact that the image set of $\sigma$ has size exactly $m$, 
imply that no two roots of $N$ have the same color under $\sigma$. 
 
\section{Tree-child networks} \label{sec:quasi-binary}
 
In this section, we shall prove that Statement~(R3) holds. To do this
we shall  reduce \leo{from} the \leo{NP-complete} \textsc{Set-Splitting} decision problem \cite[page 221]{GJ79} which is as follows. 

\begin{itemize}
	\item[$\bullet$]
 Given some \leo{finite} set $Y$, $|Y|\geq 3$, and a set $\mathcal C$ of \leo{size-3} subsets of $Y$, 
	is there a bipartition $\{A,B\}$ of $Y$ such that, for all $S \in \mathcal C$, $S \cap A \neq \emptyset$ and $S \cap B \neq \emptyset$?
\end{itemize}

 
 \begin{theorem}\label{pr-2nph}
 	For $m \geq 2$, the problem of deciding whether or not a quasi-binary tree-child $m$-network is proper forest-based is NP-complete.
 \end{theorem}
 
\begin{figure}[h]
 	\begin{center}
 		\includegraphics[scale=0.8]{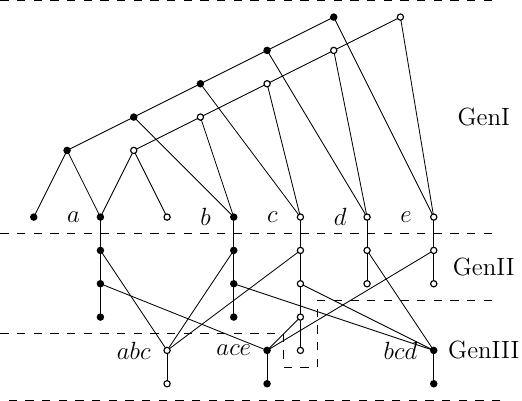}
 	\end{center}
 	\caption{For $m=2$, the $m$-network $N$ for $X=\{a,b,c,d,e\}$ 
 		and $\mathcal C=\{\{a,b,c\}, \{a,c,e\}, \{b,c,d\}\}$, as described in the proof of 
 		Theorem~\ref{pr-2nph}. Since $m=2$, there are no Gen0 vertices. GenI, GenII, and GenIII 
 		vertices are indicated as vertices in a band labelled Gen I, GenII, and GenIII, respectively. 
 		For clarity purposes, we have indicated the reticulation with the set in $\mathcal C$
 		its three parents correspond to and not the parents themselves.
 		A 2-coloring $\sigma: V(N) \to \{\bullet, \circ\}$ associated to the 
 		solution $A=\{a,b\}$, $B=\{c,d,e\}$ of the \textsc{Set-Splitting} problem for $(X,\mathcal C)$.}
 	\label{fig-ss}
 \end{figure}

 \begin{proof}
    The problem is in the class NP since, for each proper forest-based, quasi-binary, tree-child $m$-network, an $m$-coloring satisfying the conditions of Lemma~\ref{lm-fbcol} can function as a certificate and these conditions can be verified in polynomial time.
 
 	We shall prove NP-completeness by giving a reduction from \textsc{Set-Splitting}. Suppose that we are given a collection $\mathcal C$ of \leo{size-3} subsets of $X$. 
   For $m\geq 2$, we create an $m$-network 
 	as follows (see Figure~\ref{fig-ss}
 	where we illustrate the various constructions that we perform as part of this proof in terms of an example). 
 	Let $T_1$ and $T_2$ denote two isomorphic caterpillars trees on a set $Y$ with $|X|+1$ leaves. 
 	Without loss of generality we may assume that $Y=X\cup\{l\}$  and that $l\not\in X$ is a leaf 
 	in the unique cherry of $T_1$ (and therefore also of $T_2$). For all $x\in X$, we identify 
 	leaf $x$ of $T_1$ with leaf $x$ of $T_2$. The resulting network has $|X|$ reticulations. 
 	We call all vertices in the resulting DAG {\em GenI vertices}.
 	
 	If $m>2$, we add $m-1$ vertices of indegree $0$ as follows. 
 	For all $2 \leq i \leq m-1$, we add three vertices $r_i$, $h_i$, $l_i$, and 
 	the arcs $(r_i,h_i)$, $(h_i,l_i)$, and $(r_i,h_{i-1})$, where we define $h_1$ 
 	as the root of $T_2$. Then, we add two vertices $r_m,l_m$ and the arcs $(r_m,l_m)$ and $(r_m,h_{m-1})$.
 	Furthermore, we choose for all $r_i$, $2\leq i\leq m-1$, the outgoing arc $(r_i,h_{i-1})$ and subdivide it by a new vertex $s_i$. To that vertex we then add a leaf $l_i'$
 	via the arc $(s_i,l_i')$. We call all vertices added during this step {\em Gen0 vertices}.
 	
 	Assume for the remainder that $m\geq 2$. For all $x \in X$, we attach 
 	to $x$ a path $P_x$ of length $c(x)+1$ via an arc $(x,a_x)$, where $a_x$ 
 	is the first vertex on $P_x$ and  $c(x)$ is the number of sets in $\mathcal C$ containing $x$. 
 	We then bijectively label for each $x\in X$ the internal vertices of $P_x$  with the
 	elements in $\mathcal C$ that contain $x$. We call all vertices added during 
 	this step {\em GenII vertices}. Finally, for all $S \in \mathcal C$ and all $x\in S$, 
 	we create a reticulation $h_S$ with parents the vertices on $P_x$ labelled by $S$
 	and add a leaf $l_S$ to $h_S$  by adding the arc $(h_S,l_S)$. We call vertices 
 	added during this step {\em GenIII vertices}.
 	
 	Let $N$ denote the resulting DAG. By construction, $N$ is an $m$-network on $L(N)$ 
 	that is quasi-binary and tree-child. We now show that there exists a solution 
 	to the \textsc{Set-Splitting} problem for $X$ and $\mathcal C$ if and only if $N$ is proper forest-based, which 
 	will complete the proof. 
 	
 	Suppose first that $\{A,B\}$ is a solution to the \textsc{Set-Splitting} problem for $X$ 
 	and $\mathcal C$. We define an $m$-coloring $\sigma: V(N) \to \{1, \ldots, m\}$ for $N$ as follows. 
 	Let $v \in V(N)$. If $v$ is a GenI tree-vertex, we put $\sigma(v)=1$ if $v \in V(T_1)$, 
 	and $\sigma(v)=2$ if $v \in V(T_2)$. If $v$ is a Gen0 vertex, then if there exists 
 	some $2 \leq i \leq m-1$ such that $v \in \{r_i,h_i,l_i,s_i,l_i'\}$, we put $\sigma(v)=i$. 
 	Otherwise, $v\in\{r_2,h_2, l_2,r_m,l_m\}$ and we put $\sigma(v)=2$
 	if $v\in \{r_2, l_2, h_2\}$, and if $v\in \{r_m,l_m\}$ 
 	we put $\sigma(v)=m$. 
 	If $v\in V(P_x)\cup \{x\}$, some $x\in X$, then we put $\sigma(v)=1$ if $x \in A$ 
 	and $\sigma(x)=2$ if $x \in B$. Finally, if $v\in\{h_S, l_S\}$, some $S\in\mathcal C$, 
 	then we put $\sigma(v)=1$ if $|S \cap A|=1$ and $\sigma(v)=2$ if $|S \cap B|=1$. 
 	Note that since $\{A,B\}$ is a solution to the \textsc{Set-Splitting} problem, 
 	precisely one of these equalities always holds.
 	We claim that $\sigma$ satisfies (C1) and (C2) of Lemma~\ref{lm-fbcol} 
 	which implies that $N$ is proper forest-based.
 	
 	To see that (C1) holds, note first that, by construction, all 
 	non-root vertices $v$ have at least one parent $u$ satisfying $\sigma(u)=\sigma(v)$. 
 	Suppose now that $v$ is a reticulation-vertex. If $m>2$ and $v$ is a Gen0 vertex, 
 	then $v=h_i$ for some $2 \leq i \leq m-1$. In particular, $v$ is a child of $r_i$ 
 	and so $\sigma(v)=\sigma(r_i)=i$,  by definition. If $v$ is a GenI vertex, 
 	then $\sigma(v) \in \{1,2\}$ and $v$ has exactly two parents. Calling 
 	them $v_1$ and  $v_2$ we have $\sigma(v_1), \sigma(v_2) \in \{1,2\}$ 
 	and $\sigma(v_1) \neq \sigma(v_2)$. Thus, exactly one of $\sigma(v_1)=\sigma(v)$ 
 	or $\sigma(v_2)=\sigma(v)$ holds. If $v$ is a GenIII vertex, then $v$ has three parents $v_1,v_2, v_3$ and 
 	two of them must have the same color under $\sigma$. 
 	Without loss of generality assume that that $\sigma(v_2)=\sigma(v_3)$.
 	By definition, $\sigma(v)=\sigma(v_1)$ follows. Hence, (C1) holds.
 	
 	To see that (C2) holds, let $v\in V(N)-L(N)$. If $v$ is a Gen0 vertex 
 	(assuming that $N$ has such vertices!), then either $v\in\{r_i,h_i,s_i\}$, 
 	some $2\leq i\leq m-1$,  or $v\in\{r_2,h_2, r_m\}$. By definition of $\sigma$, 
 	there exists a child of $v$ that has the same color under $\sigma$ as $v$. If $v$ 
 	is a non-reticulation GenI vertex, then $v$ is an internal vertex of either $T_1$ 
 	or $T_2$. Since, for all $i=1,2$, $T_i$ is a caterpillar tree and, by definition 
 	of $\sigma$, all vertices on the directed path from the root of $T_i$ to $l$ 
 	have the same color under $\sigma$, (C2) follows. If $v$ is a GenII vertex or a 
 	reticulation GenI vertex, then $v\in \{x\}\cup V(P_x)$, some $x\in X$. 
 	Since  $P_x$ is a directed path whose first vertex is adjacent with $x$, the 
 	definition of $\sigma$ implies (C2) again.  Finally, if $v$ is a GenIII vertex, 
 	then $v=h_S$, some $S\in \mathcal C$. By definition, $\sigma(h_S)=\sigma(l_S)$. 
 	Hence (C2) also holds in this case.
 	
 	Conversely, suppose that $N$ is proper forest-based. By Lemma~\ref{lm-fbcol}, there 
 	exists an $m$-coloring $\sigma$ of $N$ satisfying (C1) and (C2). Note that, 
 	by construction, the set of all GenI reticulations  of $N$ is $X$ and also that 
 	every element in $X$  is a descendant of 
 	both $r_2$ and the root $\rho_1$ of $T_1$. By (C1), it follows that 
 	either $\sigma(x)=\sigma(r_2)$ or $\sigma(x)=\sigma(\rho_1)$ holds for all $x\in X$.  
 	Let $A=\{x\in X\,:\, \sigma(x)=\sigma(\rho_1)\}$ and $B=\{x\in X\,:\, \sigma(x)=\sigma(r_2)\}$. 
 	Clearly, $\{A,B\}$ is a bipartition of $X$ as $\sigma(r_2)\not=\sigma(\rho_1)$. 
 	
 	We claim that $\{A,B\}$ is a solution to the \textsc{Set-Splitting} problem. To see this, 
 	consider a set $S=\{x,y,z\} \in \mathcal C$. By construction,  $h_S$ has three ancestors, 
 	all of which are GenII vertices that are a descendant of $x$, $y$ and $z$, respectively. 
 	Since a GenII vertex is a vertex on $P_w$, some $w\in X$, (C1) implies that $\sigma(w)= \sigma(u)$, for all $u\in V(P_w)$.
 	Moreover, by (C2), exactly one parent $u$ of $h_S$ satisfies $\sigma(u)=\sigma(h_S)$. 
 	Without loss of generality, we may assume that $u$ is a descendant of $x$. 
 	Hence,  $\sigma(x) \neq \sigma(y)=\sigma(z)$. By definition of  $A$ and $B$, 
 	it follows that $S \cap A\neq \emptyset$ and $S \cap B \neq \emptyset$. Since 
 	this holds for all $S \in \mathcal C$, the claim follows.
 \end{proof}
 
\section{Tree-child networks revisited}
\label{sec:alternative}

In this section, we shall give an alternative proof for Statement~(R3) in the 
special case $m\ge 3$ using
colorings. We do this in part because the construction that we shall use in the proof will be 
used in the next section for establishing our results concerning binary, tree-child $m$-networks.
We shall reduce \leo{from} the \textsc{Graph $m$-colorabilty} decision problem for $m \geq 3$ (\cite[page 190]{GJ79})
which is as follows. 

\begin{itemize}
	\item[$\bullet$] Given a (simple) graph $G$, 
    does there exist a proper $m$-coloring of~$G$? 
\end{itemize}

Note that this problem can be solved in polynomial time for $m=2$ \leo{but is NP-complete for each~$m\geq 3$}. 
\leo{Hence, the reduction below} can only be used for $m\geq 3$.

\begin{proposition}\label{pr-3nph}
For $m \geq 3$, the problem of determining whether a quasi-binary tree-child 
$m$-network is proper forest-based is NP-complete.
\end{proposition}

\begin{figure}[h]
		\begin{center}
			\includegraphics[scale=0.7]{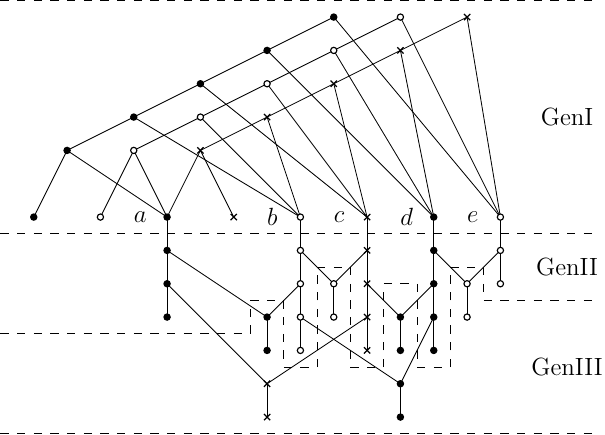}
		\end{center}
		\caption{The $3$-network $N$ constructed from the graph $G$ with vertex set $X=\{a,b,c,d,e\}$ 
			and edge set $E(G)=\{\{a,b\}, \{a,c\}, \{b,c\}, \{b,d\}, \{c,d\}, \{d,e\}\}$, as described in the proof of 
			Proposition~\ref{pr-3nph}. GenI, GenII and GenIII vertices are indicated as  described in Figure~\ref{fig-ss}.
			 A $3$-coloring $\sigma: V(N) \to \{\bullet, \circ, \times\}$ associated to the 
			 proper $3$-coloring $\kappa$ of $G$ given by $\kappa(a)=\kappa(d)=\bullet$, $\kappa(b)=\kappa(e)=\circ$ and $\kappa(c)=\times$.}
		\label{fig-3c}
\end{figure}

\begin{proof} Membership of NP can be argued in the same way as in the proof of Theorem~\ref{pr-2nph}.
We prove NP-completeness by giving a reduction from \textsc{Graph $m$-colorabilty}. 

Suppose that we are given a graph $G$ with vertex set $X$. 
Then we construct an $m$-network $N$ as follows
(see Figure~\ref{fig-3c} where we illustrate the various constructions 
performed in this proof in terms of an example). Let $T_1,\ldots, T_m$ denote $m$ 
isomorphic caterpillars trees on a set $Y$ with $|X|+1$ leaves. 
Without loss of generality we may assume that $Y=X\cup\{l\}$  and that $l\not\in X$ 
is a leaf in the unique cherry of $T_1$ (and therefore also of all $T_i$, $2\leq i\leq m$). 
For all $x\in X$ and all $1\leq i\leq m$, we identify the leaves $x$ to obtain a 
reticulation $x$ of indegree $m$. We call all vertices in the resulting graph {\em GenI vertices}.

Denoting for all $x \in X$ the degree of $x$ in $G$ by $\mathrm{deg}_G(x)$, we attach  
a directed path $P_x$ of length
$\mathrm{deg}_G(x)+1$ via an arc $(x, a_x)$ to the first vertex $a_x$ of $P_x$. We
label each internal vertex of $P_x$ with an edge in $G$ that is incident with $x$ and 
call all vertices added during this step {\em GenII vertices}. Finally, for all edges $e$ of $G$, we create a new 
reticulation $h_e$ with parents the two vertices in the DAG constructed 
thus far labelled $e$ and add a leaf $l_e$ as a child to $h_e$. We call vertices added during this step {\em GenIII vertices}.

Let $N$ denote the resulting DAG. One can easily verify that $N$ is an $m$-network that is quasi-binary and tree-child.
We now show that there exists a solution to \textsc{Graph $m$-colorabilty} for $G$ 
if and only if $N$ is proper forest-based, which will complete the proof.

Suppose first that there exists a proper $m$-coloring $\kappa: X \to \{1,\ldots, m\}$ of $G$. 
From $\kappa$, we derive  an $m$-coloring $\sigma: V(N) \to \{1,\ldots, m\}$ of $N$ as follows. 
Let $v \in V(N)$. If $v$ is a GenI tree-vertex then there exists some $1\leq i\leq m$ such that $v\in V(T_i)$. 
In this case, we put $\sigma (v)=i$. If $v$ is a GenII vertex or a reticulation GenI vertex, 
then $v\in \{x\}\cup V(P_x)$, some $x\in X$, and we put $\sigma(v)=\kappa(x)$. 
Finally, if $v$ is a GenIII vertex, then $v\in\{h_e,l_e\}$ for some 
edge $e=\{x,y\}$ of $G$. In this case, we choose $\sigma(v) \in \{\kappa(x),\kappa(y)\}$ if $v=h_e$  
and we put $\sigma(v)=\sigma(h_e)$ if $v=l_e$.

To see that $N$ is proper forest-based it suffices to show by Lemma~\ref{lm-fbcol} 
that $\sigma$ is an $m$- coloring of $N$ that satisfies Properties~(C1) and (C2). 

To see that (C1) holds, note first that, by construction, all non-root vertices $v$ of $N$ 
have at least one parent $u$ satisfying $\sigma(u)=\sigma(v)$. Suppose now that $v$ 
is a reticulation. Then $v$ is either a reticulation GenI vertex or $v=h_e$ 
some edge $e$ of $G$.  If $v$ is a reticulation GenI vertex then $v\in X$. Hence, $v$ has $m$ 
parents $v_1,\ldots, v_m$. Since, for each $1\leq i\leq m$, there exists a unique tree $T_i$ 
that contains $v_i$ it follows  that there exists a unique $1\leq j\leq m$ such that $\sigma(v)=\sigma(v_j)=j$. If $v=h_e$ then let 
$x,y\in X$ such that  $e=\{x,y\}$.
Without loss of generality, assume that $v_1$ is a vertex on $P_x$ and
that $v_2$ is a vertex on $P_y$. Assume that we have chosen $\sigma(h_e)=\kappa(x)$ 
in the definition  of $\sigma$. Then $\sigma(h_e)=\kappa(x)=\sigma(v_1)$. Since $\kappa$ 
is a proper $m$-coloring of $G$, we also have $\sigma (v_2)=\kappa(y)\not=\kappa(x)$. Thus, (C1) holds.

To see that (C2) holds, let $v\in V(N)-L(N)$. If $v$ is a non-reticulation GenI vertex, 
then $v$ has at least one child $u$ that is a non-reticulation GenI vertex. In particular, $v$
and $u$ belong to the same tree $T_i$, $1\leq i\leq m$. Hence, $\sigma(v)=\sigma(u)$, 
by the definition of $\sigma$. If $v$ is a GenII vertex or a reticulation GenI vertex, then $v$ 
has exactly one child $u$ that is a GenII vertex. By definition, $\sigma(u)=\sigma(v)$ 
also holds in this case. Finally, if $v$ is a GenIII vertex, then $v=h_e$, some edge $e$ of $G$. 
Hence, $\sigma(l_e)=\sigma(v)$ holds by definition. Thus,  (C2) holds. 

Conversely, suppose that $N$ is proper-forest-based. By Lemma~\ref{lm-fbcol}, 
there exists an $m$-coloring $\sigma: V(N) \to \{1\ldots, m\}$ of $N$ satisfying (C1) and (C2). 
Since the set of reticulation GenI vertices of $N$ is $X$, the restriction of $\sigma$ to $X$
induces an $m$-coloring $\kappa: X \to \{1,\ldots, m\}$ of~$G$. 

We claim that $\kappa$ is a proper $m$-coloring of $G$.
To see the claim, consider an edge $e=\{x,y\}$ of $G$. By construction, 
there exists a (unique) GenIII reticulation $v$ such that $v=h_e$. Let $v_1$ 
denote the parent of $h_e$ on $P_x$. Similarly, let $v_2$ denote the parent of $h_e$ on $P_y$. 
By (C1),  $\sigma(v_1)=\kappa(x)$ and $\sigma(v_2)=\kappa(y)$ hold. Since, by (C2), $\sigma(h_e)=\sigma(v_i)$ 
for a unique $i\in\{1,2\}$, say $i=1$, it follows that $\kappa(y)=\sigma(v_2)\not=\sigma(h_e)=\sigma(v_1)=\kappa(x)$. 
Thus, $\kappa$ is a proper $m$-coloring of $G$.
\end{proof}

\section{Binary tree-child networks} \label{sec:binary}

In this section, we shall prove that Statements (R1) and (R2) both hold.
We begin with some definitions. Suppose $N$ is an $m$-network and $v\in V(N)$. If $v$ is not a root of $N$, then  
we denote by $\mathcal P_N(v)$ the set of parents of $v$. Similarly, 
if $v$ is not a leaf of $N$ then we denote by $\mathcal C_N(v)$ the set of children of $v$.

\begin{lemma}\label{lem-binary}
Let $N$ be proper forest-based $m$-network, \leo{with} $m\geq 2$, and let $v$ be a 
reticulation of $N$ with indegree~$3$ or more. Then there exists a 
parent $p$ of $v$ such that the network $N'$ obtained from $N$ by:
\begin{itemize}
\item[(i)] Introducing three vertices $v_1$, $v_2$ and $x_v$.
\item[(ii)] Removing the arcs $(p,v)$ and $(v,c)$, where $c$ is the unique child of $v$ in $N$.
\item[(iii)] Adding arcs $(v,v_1)$, $(v_1,v_2)$, $(v_2,c)$, $(v_1,x_v)$ and $(p,v_2)$.
\end{itemize}
is forest-based.
\end{lemma}
\begin{proof}
Since $N$ is forest-based, Lemma~\ref{lm-fbcol} implies that  there exists an
$m$-coloring $\sigma: V(N)\to C$ of $N$ in terms of a set $C$ of colors such 
that  (C1) and (C2) are satisfied. Let $p$ be a parent of $v$ satisfying $\sigma(p) \neq \sigma(v)$ 
(which must exist in view of (C1)). Let $N'$ be the network obtained by applying Steps (i) -- (iii) to $v$ and $p$. 
Note that $N'$ also has $m$ roots.

We define an $m$-coloring $\sigma':V(N')\to C$ of $N'$ as follows. For $w \in V(N')$, we 
put $\sigma'(w)=\sigma(v)$ if $w \in \{v_1,v_2,x_v\}$, and $\sigma'(w)=\sigma(w)$ otherwise. 
Since, by construction, $V(N')=V(N) \cup \{v_1,v_2,x_v\}$ it follows that  $\sigma'$ is well-defined. 
To see that $N'$ is proper forest-based, we claim that $\sigma'$ satisfies (C1) and (C2).

To see that $\sigma'$ satisfies (C1), let $w$ be a non-root vertex of $N'$. If $w \notin \{v,v_1,v_2,x_v,c\}$,
then $w \in V(N)$ and $\mathcal P_N(w)=\mathcal P_{N'}(w)$
In particular,  $\sigma'(w)=\sigma(w)$, and $\sigma'(q)=\sigma(q)$, for all parents $q\in\mathcal P_N(w)$. 
Since, by (C1), there exists exactly one parent $q$ in $\mathcal P_N(w)$ that satisfies $\sigma(q)=\sigma(w)$ 
it follows that $q$ is the unique parent in $\mathcal P_N(w)$ satisfying $\sigma'(q)=\sigma'(w)$. 

Assume that $w \in \{v,v_1,v_2,x_v,c\}$.
If $w$ is one of $v_1,x_v$ or $c$ then $|\mathcal P_N(w)|=1 $ and $v$, $v_1$ or $v_2$ is the 
unique parent in $\mathcal P_N(w) $, respectively. By the definition of $\sigma'$, it follows 
that $\sigma'(w)=\sigma(v)$
in case $w\in\{v_1, x_v\}$. If $w=c$ then the choice of $p$ combined with the definition of $\sigma'$ 
and the fact that $\sigma$ satisfies (C1)
implies that $\sigma'(c)=\sigma(c)=\sigma(v)\not=\sigma(p)$. Similarly,
if $w=v$ then, by (C1), there exists exactly one parent $q\in \mathcal P_N(w)$ satisfying $\sigma(q)=\sigma(v)$. 
By choice of $p$, we have $p \neq q$. Since
$\mathcal P_{N'}(w)=\mathcal P_N(v)-\{p\}$
it follows that $q$ is the only parent in  $\mathcal P_{N'}(w)$ satisfying $\sigma'(w)=\sigma'(q)$.
Finally, if $w=v_2$, then 
$\mathcal P_{N'}(w)=\{v_1,p\}$. By construction, $\sigma'(w)=\sigma'(v_1)=\sigma(v)$ and, by 
choice of $p$, $\sigma'(p)=\sigma(p) \neq \sigma(v)$. Hence, $v_1$ is the only parent 
in $\mathcal P_{N'}(w)$ that has the same color as $w$ under $\sigma'$. Thus, $\sigma'$ 
satisfies (C1), as claimed.

To see that $\sigma'$ satisfies (C2), let $w$ be an internal vertex of $N'$. If $w \notin \{v,v_1,v_2,p\}$, 
then $w \in V(N)$ and $\mathcal C_N(w)=\mathcal C_{N'}(w)$.
In particular, $\sigma'(w)=\sigma(w)$ and $\sigma'(z)=\sigma(z)$, for all children $z\in\mathcal C_N(w)$. 
By (C2), there exists at least one child $z\in\mathcal C_N(w)$ satisfying $\sigma(z)=\sigma(w)$. So $\sigma'(z)=\sigma'(w)$ holds in this case.
 
Assume that $w \in \{v,v_1,v_2,p\}$.
If $w=v$, then $v_1$ is the only child in $\mathcal C_{N'}(w)$. 
Hence, $\sigma'(v_1)=\sigma(w)=\sigma'(w)$, by the definition of $\sigma'$. 
If $w=v_1$, then $\mathcal C_{N'}=\{x_v,v_2\}$. Hence, $\sigma'(w)=\sigma'(v_2)=\sigma'(x_v)=\sigma(v)$, 
by the definition of $\sigma'$. If $w=v_2$, then $\mathcal C_{N'}(w)=\{c\}$. By definition of $\sigma'$,  
we have again $\sigma'(w)=\sigma(v)$. Moreover, $\mathcal C_N(v)=\{c\}$ and $\sigma$ satisfies (C2). 
Hence, $\sigma(c)=\sigma(v)$. Since $\sigma'(c)=\sigma(c)$, we obtain $\sigma'(c)=\sigma'(w)$. 
Finally, if $w=p$, then since $\sigma$ satisfies (C2), there exists at least one child $z$ in $\mathcal C_N(p)$ 
satisfying $\sigma(z)=\sigma(p)$. Since, by choice of $p$, we have $z\not =v$
it follows that $z\in\mathcal C_{N'}(p)$. Hence, $\sigma'(z)=\sigma'(p)$, 
by the definition of $\sigma'$. Thus,  $\sigma'$  also satisfies (C2), as claimed.
\end{proof}

As a consequence of Lemma~\ref{lem-binary}, we immediately obtain 
that every quasi-binary network that is proper forest-based can be transformed into a 
binary network that is also proper forest-based. More precisely,
let $N$ be a forest-based quasi-binary $m$-network, \leo{with} $m\geq 2$, 
and let $v$ be a reticulation of $N$ of indegree $3$ or more. Then, 
in the network $N'$ obtained by applying the operations presented in the statement of 
Lemma~\ref{lem-binary} to $N$ and $v$, the indegree of $v$ in $N$ is
reduced by one  in $N'$. Moreover, by construction, all vertices $w$ of $N'$ 
such that $w \neq v$ and $w$ is a reticulation that has indegree $3$ or more in $N'$ 
is also a vertex of $N$ with the same indegree as in $N'$. In particular, the described 
operations can be recursively applied to the reticulations of $N'$ of indegree $3$ 
or more, until no such vertices remain. By Lemma~\ref{lem-binary}, the resulting network remains forest-based.

\begin{theorem}\label{pr-3nph-binary}
For $m \geq 3$, the problem of determining whether a binary tree-child $m$-network is forest-based is NP-complete.
\end{theorem}
\begin{proof} Membership of NP can again be argued in the same way as in the proof of Theorem~\ref{pr-2nph}. 
As in the proof of Proposition~\ref{pr-3nph},
we reduce from the problem of finding a proper $m$-coloring of a graph $G$ with vertex set $X$. 
    Given such a graph, we use the construction of a quasi-binary, tree-child 
    $m$-network described in the proof of Proposition~\ref{pr-3nph} to obtain a  
    quasi-binary, tree-child $m$-network~$N^s$. We then recursively apply the operations 
    in the statement of  Lemma~\ref{lem-binary} to resolve all reticulations of $N^s$ 
    with indegree 3 or more to obtain a binary, tree-child $m$-network~$N^b$. 

We next show that $N^b$ is proper forest-based if and only if $G$ admits a proper $m$-coloring. 
To this end, suppose first that $G$ admits a proper $m$-coloring. Then, by the proof of 
Proposition~\ref{pr-3nph}, $N^s$ is proper forest-based. By Lemma~\ref{lem-binary}, $N^b$ is also proper forest-based.

    Now suppose~$N^b$ is proper forest-based. Then, by Lemma~\ref{lem-binary}, 
    there exists an $m$-coloring of $N$ in terms of a set $C$ of colors that satisfies Properties~(C1) and (C2). 
    Restricting this coloring to $X$ to obtain a coloring $\kappa:X\to C$ and then applying the same arguments 
    as in the last paragraph of the proof of Proposition~\ref{pr-3nph} implies that $\kappa$ is a proper $m$-coloring of $G$. 
\end{proof}

To state the next result, we require further concepts from \cite{HMS22b}. 
Let $N$ be an $m$-network, $m\geq 2$, 
and let $v$ be a vertex of $N$. We denote by $\gamma_N(v)$ the (necessarily unique) 
\leo{lowest} ancestor of $v$ in $N$ whose indegree is not $1$. Note that $\gamma_N(v)$ is either a 
root or a reticulation of $N$. Building on this definition, we define an 
undirected graph $\Gamma(N)$ as follows. The vertex set of $\Gamma(N)$ 
is the set of all vertices  of $N$ whose indegree is not $1$. Two such 
vertices $v_1$, $v_2$ are joined by an edge in $\Gamma(N)$ if there exists 
two distinct vertices $u_1$, $u_2$ in $N$ such that $\gamma_N(u_1)=v_1$, $\gamma_N(u_2)=v_2$, 
and $u_1$ and $u_2$ share a child in $N$. \leo{The intuitive idea behind these edges is that they indicate that the vertices~$v_1,v_2$ need to be contained in different trees in a potential subdivision forest of~$N$ and so need to be assigned different colors.}

\leo{As it turns out, the edges of the graph $\Gamma(N)$ are not sufficient to determine whether~$N$ is proper forest-based. This is caused by internal vertices of~$N$ for which all children in $\mathcal C_N(v)$ have indegree $2$ or more. We call such a vertex an \emph{omnian (vertex)} \cite{jetten2016nonbinary}. To decide whether~$N$ is proper forest-based we will use}
certain supergraphs of $\Gamma(N)$ called ``omni-extensions''. 
\leo{For a semi-binary network~$N$,} we define an \emph{omni-extension} of $\Gamma(N)$
as a supergraph $\Gamma'(N)$ of $\Gamma(N)$ such that $V(\Gamma'(N))=V(\Gamma(N))$ 
and for all omnians $v$ of $N$ there exists a child $h$ of $v$ such that $\{h,\gamma_N(u)\}$ 
is an edge of $\Gamma'(N)$, \leo{with~$u$ the parent of~$h$ in~$N$ other than~$v$ \cite{HMS22a}.}
Note that $\Gamma(N)$ 
may be an omni-extension of itself. This is the case, in particular, if $N$ has no omnian vertex. If $\Gamma'(N)$ is an omni-extension of $\Gamma(N)$, 
and no proper subgraph of $\Gamma'(N)$ is an  omni-extension of $\Gamma(N)$, we say 
that $\Gamma'(N)$ is a \emph{minimal omni-extension} of $\Gamma(N)$.

\begin{figure}[h]
	\begin{center}
		\includegraphics[scale=0.7]{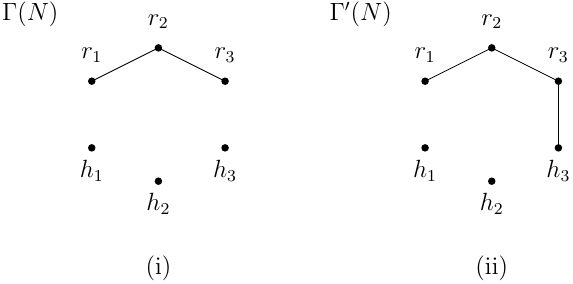}
	\end{center}
	\caption{(i) The graph $\Gamma(N)$ for the network $N$ depicted in Figure~\ref{fig:intro}. (ii) A minimal omni-extension $\Gamma'(N)$ of $\Gamma(N)$. }
	\label{fig:gamma-n}
\end{figure}

We illustrate these concepts in Figure~\ref{fig:gamma-n} \leo{for the network in Figure~\ref{fig:intro}. Observe that vertex~$v$ in Figure~\ref{fig:intro} is an omnian. At least one of the children of~$v$ needs to be assigned the same color as~$v$ (in a colouring satisfying (C2) from Lemma~\ref{lm-fbcol}). In this example, we choose child $h=h_3$. This means that the other parent~$u$ of~$h_3$ needs to be assigned a different color than~$h_3$ (by (C1)) and hence that $\gamma_N(u)=r_3$ needs to be assigned a different color than~$h_3$ (again using (C1)). This is the intuitive idea behind adding the edge $\{h_3,r_3\}$.}

\begin{theorem}
	\label{pr-binary-linear}
    Given a binary tree-child  $2$-network~$N$, it can be determined in $O(nr)$ 
    time whether~$N$ or not is proper forest-based, where~$r$ is the number of reticulations of $N$ and~$n=|L(N)|$.
\end{theorem}
\begin{proof}
    \cite[Theorem~8]{HMS22a} states that a $2$-rooted network, in which all reticulations 
    have indegree~2, is proper forest-based if and only if the graph~$\Gamma(N)$ has a 
    bipartite omni-extension. This theorem is applicable to~$N$ since it is binary. 
    Since~$N$ is tree-child, it has no omnians. Hence,~$\Gamma(N)$ is an omni-extension of~$\Gamma(N)$.

    We claim that~$N$ is proper forest-based if and only if~$\Gamma(N)$ is bipartite. 
    First suppose that~$N$ is proper forest-based. Then, by \cite[Theorem~8]{HMS22a} recalled above, 
    $\Gamma(N)$ has a bipartite omni-extension. Since~$\Gamma(N)$ is a subgraph of every omni-extension 
    of $\Gamma(N)$, it follows that~$\Gamma(N)$ is bipartite. Conversely, if~$\Gamma(N)$ is bipartite, 
    then, since~$\Gamma(N)$ is an omni-extension of $\Gamma(N)$, it follows again by \cite[Theorem~8]{HMS22a} 
    recalled above that~$N$ is proper forest-based. The claim therefore holds.

    To construct~$\Gamma(N)$, we need to find, for each parent $v$ of a reticulation of $N$, 
    the vertex $\gamma_N(v)$. This takes $O(|V(N)|)$ time per reticulation of $N$. 
    Hence, the construction of~$\Gamma(N)$ takes $O(|V(N)|\cdot r)$ time.
    Checking whether~$\Gamma(N)$ is bipartite or not takes $O(r)$ time, since~$\Gamma(N)$ 
    has at most~$r$ edges. Hence, the total running time is $O(|V(N)|\cdot r)$.
    
    Finally, to obtain the running time stated in the theorem, note that~$N$ has $O(n)$ 
    vertices, since by~\cite[Proposition~10.7]{S16}, a tree-child $1$-network has 
    fewer than~$4n$ vertices. Adding a root $\rho$ and an arc from $\rho$ to each of the 
    original roots of a  tree-child $2$-network preserves the tree-child property and 
    increases the number of vertices by only~$1$. This concludes the proof.
\end{proof}

\section{An FPT algorithm}
\label{sec:ftp-alg}

In this section, we establish Statement~(R4).
We shall present a fixed-parameter tractable (FPT) algorithm called {\sc Check} 
for deciding whether a semi-binary $m$-network $N$, $m\geq 2$, is forest-based, 
with parameters the number~$r$ of reticulations of~$N$, the number~$m$ of roots of~$N$ \leo{and the maximum outdegree~$\Delta$ of~$N$}. 
Note that we do not require the network to be tree-child. 

\begin{algorithm}[h!]
	\caption{\label{alg:simplification-seq} The algorithm {\sc Check}.}
	\begin{algorithmic}[1]
		\Require{Semi-binary $m$-network $N$, some $m\geq 2$.}
		\Ensure{The statement ``$N$ is proper forest-based'' or the statement ``$N$ is not proper forest-based''.}
		\State Construct $\Gamma(N)$ 
		and find all minimal omni-extensions of $\Gamma(N)$.
		 \ForAll{minimal omni-extension of $\Gamma(N)$}
		Add an edge between any two distinct roots of~$N$ to obtain a graph $\Gamma^*(N)$. Note that $\Gamma^*(N)$ is also an omni-extension of $\Gamma(N)$.
		\State Find all proper $m$-colorings of $\Gamma^*(N)$.
		\ForAll{proper $m$-coloring $\sigma$ of $\Gamma^*(N)$} 
		\ForAll{reticulations~$h$ of $N$ and all roots~$\rho$ of $N$} check that if
		$\sigma(h)=\sigma(\rho)$ then  there exists a directed path $P(\rho,h)$ from~$\rho$ to~$h$ in~$N$ such that $\sigma(h)=\sigma(h')$, 
		for all reticulations $h'$ of $N$ on~$P(\rho,h)$. 
		\EndFor
		\If{$P(\rho,h)$ exists for all reticulations $h$ of $N$ and all
			roots $\rho$ of $N$ with $\sigma(h)=\sigma(\rho)$} return ``$N$ is proper forest-based''.
		\EndIf
		\EndFor
		\EndFor
		\If{for each minimal omni-extension of $\Gamma(N)$ there is no proper $m$-coloring of $\Gamma^*(N)$ with the required paths} return ``$N$ is not proper forest-based''.
		\EndIf
		\label{alg:alg1-last-line}
	\end{algorithmic}
\end{algorithm}






Correctness of the algorithm {\sc Check} follows from Theorem~\ref{thm:minimal}
which is a slight strengthening of \cite[Theorem 7]{HMS22a} 
to minimal omni-extensions. For $N$ an $m$-network, $m\geq 2$, we denote the set of roots and reticulations of $N$ by $RH(N)$.

\begin{theorem}\label{thm:minimal} 
	Let $N$ be a semi-binary $m$-network, some $m\geq 2$, and let
	$\{s_1,\ldots, s_m\}$ be a set of $m$ colors. Then $N$ is a proper 
	forest-based if and only if there exists a minimal omni-extension $\Gamma'(N)$ of $\Gamma(N)$ 
	and a proper $m$-coloring $\sigma: RH(N) \to \{s_1, \ldots, s_m\}$ of $\Gamma'(N)$ satisfying:
	\begin{itemize}
		\item[(F1)] The restriction of $\sigma$ to the set $R(N)$ of roots of $N$ is a bijection.
		\item[(F2)] For all $u \in R(N)$ and all reticulations $ v$ of $N$
		such that $\sigma(u)=\sigma(v)$ there
		exists a directed path $P$ in $N$ from $u$ to $v$ such that $\sigma(w)=\sigma(u)$ 
		holds for all reticulations $w$ of $N$ that lie on $P$.
	\end{itemize}
\end{theorem}

\begin{proof}
  Suppose first that there exists a minimal omni-extension $G$ of $\Gamma(N)$ and a proper $m$-coloring of $G$ satisfying Properties~(F1) and (F2). Then by \cite[Theorem 7]{HMS22a}, $N$ is proper forest-based.

 Conversely, suppose that $N$ is proper forest-based. Then by \cite[Theorem 7]{HMS22a}, there exists an omni-extension 
	$G$ of $\Gamma(N)$ and a proper $m$-coloring $\sigma$ of $G$ satisfying Properties~(F1) and (F2). Since~(F1) and (F2) 
		are independent of the structure of $G$, it follows 
	for all subgraphs $G^-$ of $G$ with 
	$V(G)=V(G^-)$ that $\sigma$ is a proper $m$-coloring of $G^-$ satisfying (F1) and (F2). In particular, if there exists an omni-extension of $\Gamma(N)$ 
	and a proper $m$-coloring of $G$ satisfying (F1) and (F2), then there exists a minimal omni-extension 
	of $\Gamma(N)$ and a proper $m$-coloring of $G$ satisfying Properties~(F1) and (F2).
\end{proof}

 We now analyze the run time of algorithm {\sc Check}. 
 \leo{The graph $\Gamma(N)$ can be constructed in $O(r|V(N)|)$ time and} can have at most \leo{$\Delta^{\omega}$} 
 minimal omni-extensions where~$\omega$ is the number of omnians of $N$.
 \leo{Since~$N$ is semi-binary, $\omega$ is at most~$2r$ (each omnian is the parent of at least one reticulation and each reticulation has two parents).}
 All proper $m$-colorings of $\Gamma^*(N)$ can be found in $O(m^{r+m}\leo{(r+m)^2})$ time since $\Gamma^*(N)$ has $r+m$ vertices (Line 3).
 Since, by construction, the set of roots on $N$ forms a clique in
 $\Gamma^*(N)$ and we are interested in proper $m$-colorings of 
 $\Gamma^*(N)$ it follows that for every reticulation  $h$ of $N$ there exist a 
 unique root $\rho_h$ of $N$ that has the same color under the  $m$-coloring under consideration as $h$. 
 In the worst case, checking the vertices on the directed path $P(\rho,h)$
 takes $O(|V(N)|)$ time per pair $(\rho_h,h)$ (by deleting all reticulations that do not have the 
 same color as~$\rho$ and then doing a depth-first search) (Line 5). The total run time 
 of algorithm {\sc Check} therefore is $O(\leo{\Delta^{2r}}m^{r+m}\leo{(r+m)^2}|V(N)|)$.

\begin{theorem} 
	\label{theo:ftp-alg}For all $m\geq 2$, there exists an algorithm with 
	running time \leo{$O(\Delta^{2r}m^{r+m}(r+m)^2v)$} to decide whether a semi-binary $m$-network with~$r$ 
	reticulations,~$v$ vertices \leo{and maximum outdegree~$\Delta$} is proper 
	forest-based.
\end{theorem}

\section{Discussion}
\label{sec:conclusion}

We have shown that it can be decided in polynomial time whether or not a binary, tree-child $m$-network $N$ is proper forest-based for $m=2$, but that this problem is NP-complete for $m\ge 3$. It would be interesting to know if the same problem can be solved in polynomial time in case $m=2$ but $N$ is not necessarily binary, for example, in case $N$ is tree-sibling (i.e. every reticulation has at least one sibling vertex that is a tree vertex \leo{or a leaf}), or $N$ is an arbitrary 2-network. 
In addition,  although we have shown that there is an FPT algorithm for 
deciding whether or not a semi-binary $m$-network is forest-based, it would be interesting to see if FPT-algorithms with improved run-times can be developed, 
or if an FTP algorithm can be found for arbitrary $m$-networks.

In this paper, we have not considered the problem of deciding whether or not an $m$-network is forest-based, $m\ge 1$, i.e. the problem where we do not insist that the underlying forest must have $m$ trees. 
 It would be interesting to know whether or not this is an NP-complete problem. 
Note that since an $m$-network $N$ is forest-based if and only if it contains a subdivision forest in which every tree is an induced, directed path in $N$ 
 (cf. \cite[Theorem 1]{HMS22a}), this question is closely related to the 
 induced path partition problem which is known to NP-complete \cite{fernau2023parameterizing}.

Finally, for biological applications, it would be useful to develop new algorithms to construct forest-based networks from biological data.
Hopefully the structural insights that we have developed in this paper for proper forest-based networks will provide some helpful insights for this problem.

 \medskip

\noindent{\bf Acknowledgement}
The authors would like to thank the Institute Mittag Leffler, Sweden, for hosting the ``Emerging Mathematical Frontiers in Molecular Evolution" meeting where the ideas underlying this paper were conceived. LvI would like to thank the Netherlands Organization for Scientific Research (NWO) grant OCENW.KLEIN.125.

\bibliographystyle{plain}
\bibliography{alpha}

\end{document}